\newcommand*{\ATLASLATEXPATH}{}
\documentclass[cernpreprint, atlasdraft=false, texlive=2016, USenglish]{\ATLASLATEXPATH atlasdoc}
 
\usepackage{\ATLASLATEXPATH atlaspackage}
\usepackage{rotating}
\usepackage{\ATLASLATEXPATH atlasbiblatex}
 
\usepackage{\ATLASLATEXPATH atlasphysics}
 
\graphicspath{{logos/}{figures/}}
 
\usepackage{ANA-HIGG-2019-01-PAPER-defs}

\AtlasTitle{$CP$ Properties of Higgs Boson Interactions with Top Quarks in the $t\bar{t}H$ and $tH$ Processes Using $H \rightarrow \gamma\gamma$ with the ATLAS Detector}
 
\AtlasAbstract{ A study of the $CP$ properties of the interaction between the Higgs boson and
top quarks is presented. Higgs bosons are identified via the
diphoton decay channel ($H \rightarrow \gamma\gamma$) and their production in association with
a top quark pair ($t\bar{t}H$) or single top quark ($tH$) is studied.
The analysis uses 139 fb$^{-1}$ of proton--proton collision
data recorded at a center-of-mass energy of $\sqrt{s}$ = 13 TeV with the
ATLAS detector at the Large Hadron Collider.
Assuming a $CP$-even coupling, the $t\bar{t}H$ process is observed with
a significance of 5.2 standard deviations. The measured
cross section times $H \to \gamma \gamma$ branching ratio is
$1.64\;^{+0.38}_{-0.36}\;\text{(stat.)}\;^{+0.17}_{-0.14}\;(\text{sys.})$ fb,
and the measured rate for $t\bar{t}H$ is
$1.43\;^{+0.33}_{-0.31}\;\text{(stat.)}\;^{+0.21}_{-0.15}\;(\text{sys.})$ times the Standard Model
expectation.
The $tH$ production process is not observed and an upper limit on its
rate of 12 times the Standard Model expectation is set.
A $CP$-mixing angle greater (less) than 43 (-43)$^\circ$ is excluded at 95$\%$ confidence level.
}
 
\author{The ATLAS Collaboration}
 
\AtlasRefCode{HIGG-2019-01}
 
\AtlasJournalRef{Phys. Rev. Lett. 125 (2020) 061802}
\AtlasDOI{10.1103/PhysRevLett.125.061802}
 
\PreprintIdNumber{CERN-EP-2020-046}
 
\hypersetup{pdftitle={ATLAS document},pdfauthor={The ATLAS Collaboration}}
 
\begin{document}
 
\maketitle

The observation of Higgs boson production in association with top quarks at the LHC~\cite{HIGG-2018-13, CMS-HIG-17-035}
provides an opportunity to probe the charge conjugation and parity~($CP$) properties of the
Yukawa coupling of the Higgs boson to the top quark.
The Standard Model (SM) of particle physics predicts the Higgs boson to be a scalar
particle ($J^{CP}$~=~0$^{++}$) with a prescribed coupling to the top quark. However, the presence of a $J^{CP}$~=~0$^{+-}$ pseudoscalar admixture, which introduces a second coupling to the top quark,
has not yet been excluded.
Any measured $CP$-odd contribution would be a sign of physics beyond the SM
and could account for the explanation of the observed baryon asymmetry of the universe.
This Letter presents a search for $CP$-violation in this coupling and measurements of
the production rate of the Higgs boson, via its decay into two photons, in association with top quarks.
Recently, the CMS Collaboration performed a similar study~\cite{CMS-TTH-20-028}.
 
Studies of $CP$ properties of the Higgs boson interactions  with gauge bosons have been performed
by the ATLAS and CMS experiments~\cite{HIGG-2015-06, HIGG-2016-22, HIGG-2016-21,
CMS-HIG-14-035, CMS-HIG-18-002, CMS-HIG-17-034}; the results show no deviations from the SM predictions.
However, these measurements probe
the bosonic couplings in which $CP$-odd contributions enter only via higher-order operators
that are suppressed by powers of 1/$\Lambda^2$~\cite{efttop:2011prd}, where $\Lambda$ is the scale of the
new physics in  an effective field theory~(EFT).
In the case of the Yukawa couplings,
the $CP$-odd contributions are not suppressed by powers of 1/$\Lambda^2$.

The $CP$ properties of the top Yukawa
coupling can be probed directly using Higgs boson production in association
with top quarks: $t\bar{t}H$ and $tH$ processes.
The couplings
impact the production
rates~\cite{Ellis:2013yxa, Boudjema:2015nda, Buckley:2016prl, AmorDosSantos:2017ayi} and
some kinematic distributions. The $tH$ rate is particularly sensitive to deviations from SM
couplings due to destructive interference in the SM between diagrams where the Higgs boson radiates from a top quark and
from a $W$ boson.
The presence of $CP$-mixing
in the top Yukawa coupling also modifies the gluon--gluon
fusion~(ggF) production rate and the $H\to\gamma\gamma$ decay rate. 
 
\newcommand{\AtlasCoordFootnote}{
ATLAS uses a right-handed
coordinate system with its origin at the nominal interaction point
(IP) in the center of the detector and the \(z\)-axis along the beam
pipe.  The \(x\)-axis points from the IP to the center of the LHC
ring, and the \(y\)-axis points upwards.  Cylindrical coordinates
\((r,\phi)\) are used in the transverse plane, \(\phi\) being the
azimuthal angle around the \(z\)-axis.  The pseudorapidity is
defined in terms of the polar angle \(\theta\) as \(\eta = -\ln
\tan(\theta/2)\).
Angular distance is measured in units of \(\Delta   R \equiv \sqrt{(\Delta\eta)^{2} + (\Delta\phi)^{2}}\).
}
 
This analysis is performed using 139~fb$^{-1}$ of $\sqrt{s}$~=~13~TeV proton--proton~($pp$)
collision data recorded from 2015 to 2018 with the ATLAS detector.
The ATLAS detector~\cite{PERF-2007-01, ATLAS-TDR-2010-19, PIX-2018-001} is a multipurpose particle
detector with a forward--backward symmetric cylindrical geometry
and near $4\pi$ coverage in solid angle.\footnote{\AtlasCoordFootnote}
The trigger system consists of a hardware-based first-level trigger and a software-based
high-level trigger~\cite{TRIG-2016-01}.
Events used in this analysis were triggered by requiring two photons
with a \textit{loose} identification requirement~\cite{EGAM-2018-01}
in the 2015--2016 data-taking period and transverse energies of at least 25~GeV and
35~GeV for the subleading and leading photons, respectively.
Due to the greater instantaneous luminosity,
the photon trigger identification
requirement was tightened
in the 2017--2018 data-taking period.
The average trigger efficiency is over $98\%$ for events passing the full diphoton
event selection for this analysis.
 
The EFT definition used in this Letter is provided by the Higgs
Characterization model~\cite{Demartin:2014fia}, which is implemented
in the \textsc{MadGraph5\_aMC@NLO} generator~\cite{Alwall:2014hca}.  Within this model,
the term in the effective Lagrangian that describes the top Yukawa coupling
is:
\begin{align*}
\cal{L} = &-\frac{m_t}{v} \left\{ \bar{\psi_{t}}\kappa_t \left[\cos(\alpha) + \text{i}\sin(\alpha)\gamma_{\text{5}} \right]\psi_{t} \right\} H
\end{align*}
 
where $m_{t}$ is the top quark mass, $v$ is the Higgs vacuum expectation value,
$\kappa_t$~($>0$) is the top Yukawa coupling parameter,
and $\alpha$ is the $CP$-mixing angle.
The SM corresponds to a $CP$-even coupling with $\alpha=0$ and $\kappa_t=1$
while a $CP$-odd coupling is realized when $\alpha=90^\circ$.
 
Simulated $t\bar{t}H$ and $tH$ samples were generated using \MGMCatNLOV{2.6.2}
at next-to-leading order in QCD for different
$\alpha$ and $\kappa_t$~(for $tH$) values,
with the NNPDF30NLO~\cite{Ball:2014uwa} parton distribution function~(PDF) set
used for the matrix element~(ME) evaluation,
and interfaced to \PYTHIAV{8}~\cite{Pythia8} using the NNPDF23LO~\cite{Ball:2012cx} PDF set
for parton showering~(PS).
The A14 parameter set~\cite{ATL-PHYS-PUB-2014-021}, tuned to data, was
used for both PS and underlying event (UE).
From these samples, the yields for $t\bar{t}H$ and $tH$ are parameterized as functions of
$\alpha$ and $\kappa_t$, which are used in the statistical interpretations.
Samples for other Higgs boson production processes, ggF~\cite{Hamilton:2013fea},
vector-boson fusion (VBF)~\cite{Nason:2010jhep},
and vector-boson associated production ($VH$)~\cite{Mimasu:2015nqa,Luisoni:2013kna}
were produced with \POWHEGBOXV{v2} generator~\cite{Powheg_Alioli}
using the PDF4LHC15 PDF set~\cite{Butterworth:2015oua} for ME,
with the AZNLO set of tuned parameters~\cite{STDM-2012-23} and
\PYTHIAV{8} for PS using the CTEQ6L1~\cite{Pumplin:2002vw} PDF set.
Samples generated with \HerwigV{7}~\cite{Bellm:2015jjp} are used for
systematic uncertainty studies that
involve modeling of
the PS, hadronization and UE.
The simulated Higgs boson samples are normalized to the SM cross sections (Refs.~\cite{YR4,Anastasiou:2015ema, Anastasiou:2016cez, Actis:2008ug, Anastasiou:2008tj,Ciccolini:2007jr,Ciccolini:2007ec,Bolzoni:2010xr,Brein:2003wg,Altenkamp:2012sx,Denner:2011id,Beenakker:2002nc,Dawson:2003zu,Yu:2014cka,Frixione:2015zaa,Dawson:2003kb,Dittmaier:2003ej,Harlander:2011aa,Demartin:2015uha}) times the SM branching ratio~(BR) to diphotons (Refs.~\cite{YR4,Djouadi:1997yw,Djouadi:2006bz,Bredenstein:2006ha,Bredenstein:2006rh}) with a Higgs boson mass of 125.09 GeV~\cite{HIGG-2014-14}, and specifically for $t\bar{t}H$ the SM predicted
cross section times the $H \to \gamma \gamma$ BR is
$\sigma_{t\bar{t}H}\times B_{\gamma\gamma} = 1.15\;_{-0.12}^{+0.09}$ fb.
 
Although this analysis relies on a data-driven approach for background estimations,
a simulated background sample for the $t\bar{t}\gamma\gamma$ process was generated
to optimize the event selection and develop the background model.
This sample was generated using the
\textsc{Madgraph5\_aMC@NLO} generator, with the NNPDF23LO PDF set
and showered with \PYTHIAV{8}.
 
All generated Higgs boson events were passed through a full simulation of the
ATLAS detector response~\cite{SOFT-2010-01} using \GEANT4~\cite{Geant4}. The
$t\bar{t}\gamma\gamma$ events were processed with a fast simulation
in which the full simulation of the calorimeter is replaced with a parameterization of the
calorimeter response~\cite{ATL-PHYS-PUB-2010-013}. The effects of multiple $pp$ interactions
in the same or neighboring bunch crossings are included using events generated with \PYTHIAV{8}.
Events are weighted such that the distribution of the average number of interactions per bunch
crossing matches that observed in data, which is typically around 30 -- 40.
 
Events are selected by requiring two isolated photon candidates with
transverse momenta $\pt$ greater than 35~GeV and 25~GeV.
Both photons must satisfy the \textit{tight} identification requirement~\cite{EGAM-2018-01}.
The identification is constructed from a cut-based selection
using the electromagnetic shower shape variables.
The leading (subleading) photon must have $\pt/\myy$ > 0.35 (0.25), and
the diphoton invariant mass \myy\ is required to be in the range $\myy \in [105, 160]\,\GeV$.
Jets are reconstructed using the anti-$k_{t}$ algorithm~\cite{Cacciari:2008gp}
with a radius parameter of $R=0.4$.
Events are required to have at least one jet with $\pt > 25$~\GeV\  containing a $b$-hadron~($b$-jet),
identified using a $b$-tagging algorithm with an efficiency of 77\% and a mistagging rate of 0.9\%
for light-flavor jets~\cite{FTAG-2018-01}.
 
Selected events are sorted into two $t\bar{t}H$-enriched regions.
The ``Lep'' region, targeting top quark  decays in which at least one of the resulting
$W$ bosons decays leptonically, requires events to have at least one
isolated lepton (muon or electron) candidate with $\pt > 15$~\GeV\  passing
\textit{medium} identification requirements~(Refs.~\cite{EGAM-2018-01, PERF-2015-10}).
The ``Had'' region targets hadronic top quark
decays (as well as top quark decays to both hadronically decaying $\tau$-leptons
and unreconstructed leptons) and requires events to have at least two additional
jets with $\pT>25$ GeV and no selected lepton.
 
A boosted decision tree (BDT) used for the top quark reconstruction,
denoted by ``Top Reco BDT'', is trained with the $t\bar{t}H$ sample by using the \textsc{XGBoost}
package~\cite{XGBoost}, to extract the three-jet (triplet) combination
best matching the hadronic decay products of a top quark.
This BDT uses \pt, $\eta$, $\phi$,
and the energy $E$ of $W$ boson and $b$ jet~(where the $W$
boson candidate is formed by a pair of jets).
Furthermore, this BDT uses
the angular distance $\Delta R_{Wb}$ between the $W$ boson and $b$ jet,
$\Delta R_{jj}$ between the two jets composing the $W$ boson candidate, and $b$-tagging
information about all three jets in the triplet and the invariant mass of the triplet.
For events in the Had region, the triplet with the highest Top Reco BDT score
is taken as the primary top quark candidate ($t_{1}$). In the Lep region, for events containing
only one lepton, a $W$ boson candidate is first constructed from
the lepton and missing transverse momentum $E_\text{T}^\text{miss}$.
Then $t_{1}$ is reconstructed from this
leptonic $W$ boson candidate and the jet giving the highest
Top Reco BDT score. No top quark candidate is reconstructed for events containing more than one lepton.
After $t_{1}$ is selected,
if there are at least three additional jets,
a second top quark candidate~($t_{2}$) is reconstructed by selecting
the triplet with the highest BDT score
from the remaining jets; if there is only one or two additional jets,
then $t_{2}$ is taken as the sum of
the remaining jets; otherwise no $t_{2}$ is reconstructed.
 
To improve the analysis sensitivity, selected events are categorized using partitions
of a two-dimensional BDT space. Two independent BDTs are trained using the \textsc{XGBoost}
algorithm: ``Background Rejection BDT'' and ``$CP$ BDT'', and each of them is
trained separately in the Had and Lep regions.
The Background Rejection BDT is trained to separate $t\bar{t}H$-like events from background
which are mainly non-resonant diphoton production processes
including $\gamma\gamma$+jets and $t\bar{t}\gamma\gamma$, and
a detailed discussion of  this methodology is given in Ref.~\cite{HIGG-2018-13}.
The $CP$ BDT is trained to separate $CP$-even from $CP$-odd couplings 
using $t\bar{t}H$ and $tH$ processes. The $CP$ BDT uses
$\pt$ and $\eta$ of the diphoton system,
\pt and $\eta$ of $t_{1}$ and $t_{2}$,
their azimuthal angles calculated relative to the diphoton system
$\phi_{\gamma \gamma, t_{1}}$, $\phi_{\gamma \gamma, t_{2}}$,
as well as their Top Reco BDT scores.
It also uses differences in pseudorapidity and azimuthal angle
$\Delta \eta_{t_{1} t_{2}}$ and $\Delta \phi_{t_{1} t_{2}}$ between the two top quark candidates,
the invariant mass of the diphoton and primary top quark system $m_{\gamma \gamma, t_{1}}$,
the invariant mass of the two top quark candidates $m_{t_{1} t_{2}}$, the scalar \pt sum of jets $H_\text{T}$,
the $E_\text{T}^\text{miss}$ divided by $\sqrt{H_\text{T}}$, the number of jets and $b$-tagged
jets, and the minimum and second-smallest angular differences $\Delta R_{\gamma j}$ between
a photon and a jet.
 
\begin{figure}[htbp]
\centering
\includegraphics[width=0.8\textwidth]{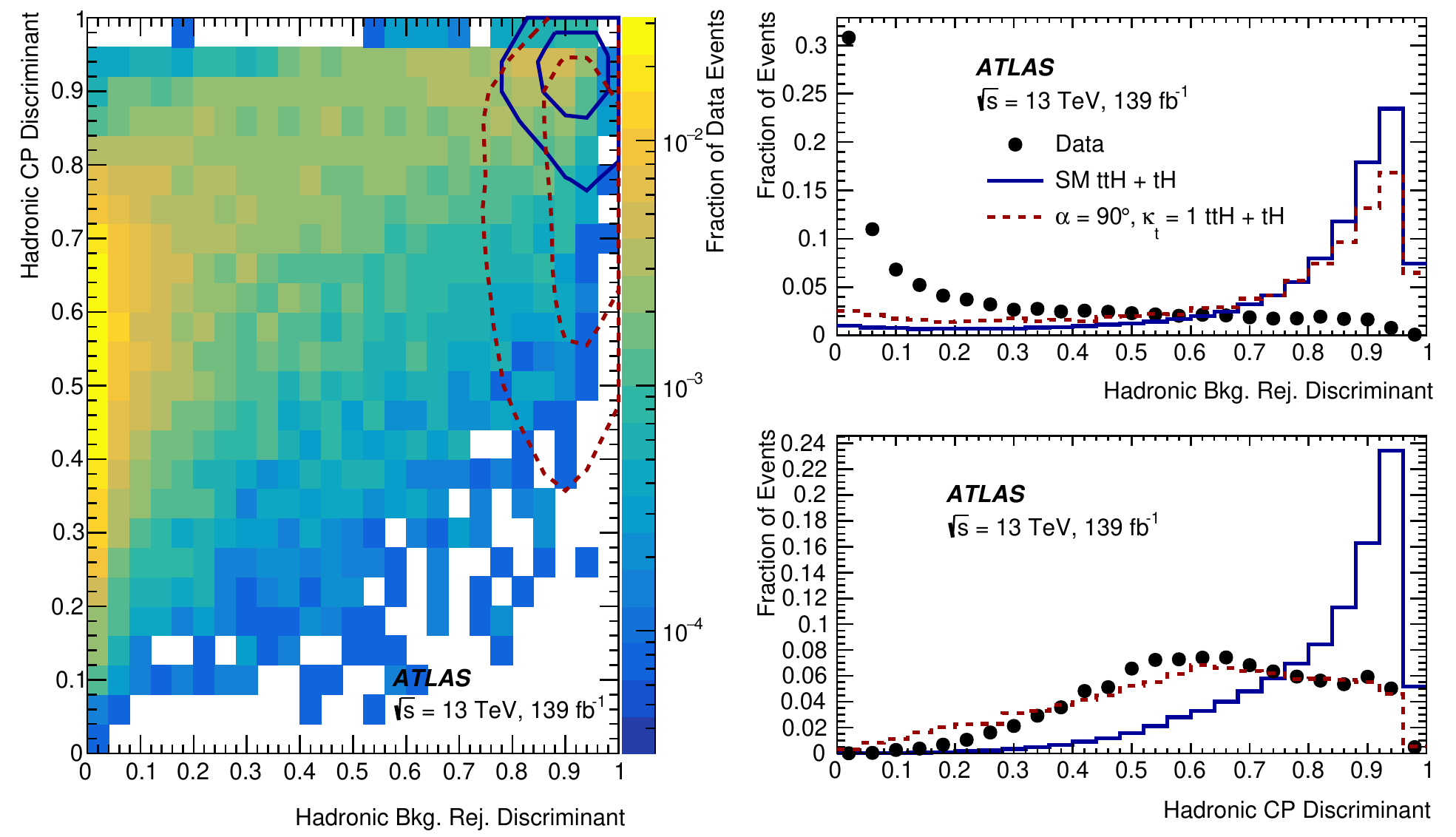}
\caption{Left: two-dimensional BDT distribution in the selected data events
($\myy \in [105, 160]\,\GeV$) from the Had region
showing the Background Rejection BDT and
$CP$ BDT. The inner (outer) contours capture 25\% (50\%) of the $t\bar{t}H$ and $tH$ signal events
for $CP$-even (blue) and $CP$-odd (red) hypotheses.
Right: projections onto the background rejection and $CP$ BDT axes.
Contributions from $CP$-even (blue), $CP$-odd (red) $t\bar{t}H$/$tH$ processes and
the data (black) are shown
and normalized to unit area.
The error bars on data are statistical.}
\label{fig:bdt:contours}
\end{figure}

Figure~\ref{fig:bdt:contours}~(right)
shows the BDT discriminant distributions in the data
as well as those expected from $CP$-even and $CP$-odd Higgs boson signals in the Had region.
The discriminating power can be seen by comparing the $CP$-even, $CP$-odd and data shapes.
Events with low values of the Background Rejection BDT response are removed and
the remaining events are categorized.
The number of
categories and the boundary locations
are chosen to optimize  the $t\bar{t}H$ significance and the discriminating power
between the $CP$-even and $CP$-odd cases. 
There are 20 categories in total, 12 in the Had region
and 8 in the Lep region.

The results are impacted by three distinct types of uncertainties: the statistical uncertainty associated with the data, theoretical modeling systematic uncertainties
and experimental systematic uncertainties. The first dominates.
Theoretical uncertainties for  $t\bar{t}H$ and $tH$ rates in the various categories are assessed.
The following effects are considered: the value
of the strong coupling constant;
alternative generator for the PS, hadronization and UE; PDF uncertainty.
In the three (two) most $CP$-even sensitive Had (Lep) categories,
each of these uncertainties is less than 10\%.
The background from ggF is
less than 0.25 events in each of the most sensitive categories;
conservative uncertainties, including a 100\% theoretical uncertainty
in the modeling of the radiation of additional heavy-flavor jets,
are assigned to it in the Had region.
The same heavy-flavor uncertainty is also assigned to the VBF and $VH$ processes.

Experimental uncertainties arise from identification and isolation
criteria for photons, electrons and muons and from their energy
scale and resolution~\cite{EGAM-2018-01, PERF-2015-10}.
Jets have uncertainties from $b$-tagging~\cite{FTAG-2018-01}
and vertex identification~\cite{PERF-2014-03}
in addition to the energy scale and resolution~\cite{PERF-2016-04}. Uncertainties in the measurement of
$E_\text{T}^\text{miss}$~\cite{PERF-2016-07},
which is used in the leptonic categories, are also included.
These experimental effects impact the expected event yield in each category and can cause events to migrate between the categories.
The overall uncertainty is less than 20\% in each category.
In addition, uncertainties in the luminosity~\cite{ATLAS-CONF-2019-021}
obtained using the LUCID-2 detector~\cite{LUCID2} and
trigger efficiency~\cite{TRIG-2016-01} are responsible for uncertainties in the overall event yield of 1.7\% and 0.4\%, respectively.
 
A simultaneous maximum-likelihood fit is performed to the
$m_{\gamma\gamma}$ spectra in all the categories.
Signal and background shapes are modeled by analytic functions using the strategy
discussed in Ref.~\cite{HIGG-2016-21}.
The chosen background function is based on the simulated
$t\bar{t}\gamma\gamma$ events following the procedure in Ref.~\cite{HIGG-2018-13},
which imposes stringent conditions on potential biases in
the extracted signal yield to avoid losses in sensitivity.
The parameters of the background model and background normalization
in each category are left free in the fit.
The profile likelihood ratio is used as the test statistic and the
asymptotic approximation~\cite{Cowan:2010js} is used for statistical interpretations.
Yields from $\ttH$ and $tH$ are extracted after
subtracting the very small contribution from other Higgs boson production modes
using their SM expected values.
Figure~\ref{fig:mggmt_mc} shows the distributions of the reconstructed masses for
the diphoton system and primary top quark.
The events
are weighted by $\ln(1 + S/B)$ with $S$ and $B$ being the fitted signal and background
yields in the smallest \myy\ interval containing 90\% of the signal in each category.
The $p$ value associated with the compatibility between the observed spectra
and the fit model using the
goodness-of-fit test
method described in Ref.~\cite{Goodnessfit} is 35\%.
Assuming a $CP$-even coupling, the $\sigma_{t\bar{t}H}\times B_{\gamma\gamma}$ is derived by
constraining all the non-$t\bar{t}H$ Higgs boson processes to their SM predictions
and measured to be $1.64\;^{+0.38}_{-0.36}\;\text{(stat.)}\;^{+0.17}_{-0.14}\;(\text{sys.})$ fb.
The measured rate for $t\bar{t}H$ is
$1.43\;^{+0.33}_{-0.31}\;\text{(stat.)}\;^{+0.21}_{-0.15}\;(\text{sys.})$ times the SM expectation.
The background-only hypothesis is rejected with an observed (expected)
significance of $5.2\sigma$ ($4.4\sigma$).
The rate for $tH$ is derived by constraining all the
non-$t\bar{t}H$/$tH$ Higgs boson processes to their SM prediction
without prior constraint on the rate of $t\bar{t}H$.
Using the CLs method~\cite{Read:2002hq}, this yields a
95\% confidence level~(CL) upper limit of 12 times the SM prediction, the same as expected
assuming the presence of SM $tH$ signal.
This is stricter than the previous best limit of 25 times the SM prediction
on $tH$ from the CMS analysis performed using 35.9~\ifb\ of
data at $\sqrt{s}$~=~13~TeV~\cite{CMS-HIG-18-009}
with the $t\bar{t}H$ process constrained to the SM prediction.
 
\begin{figure}[htbp]
\centering
\includegraphics[width=0.8\textwidth]{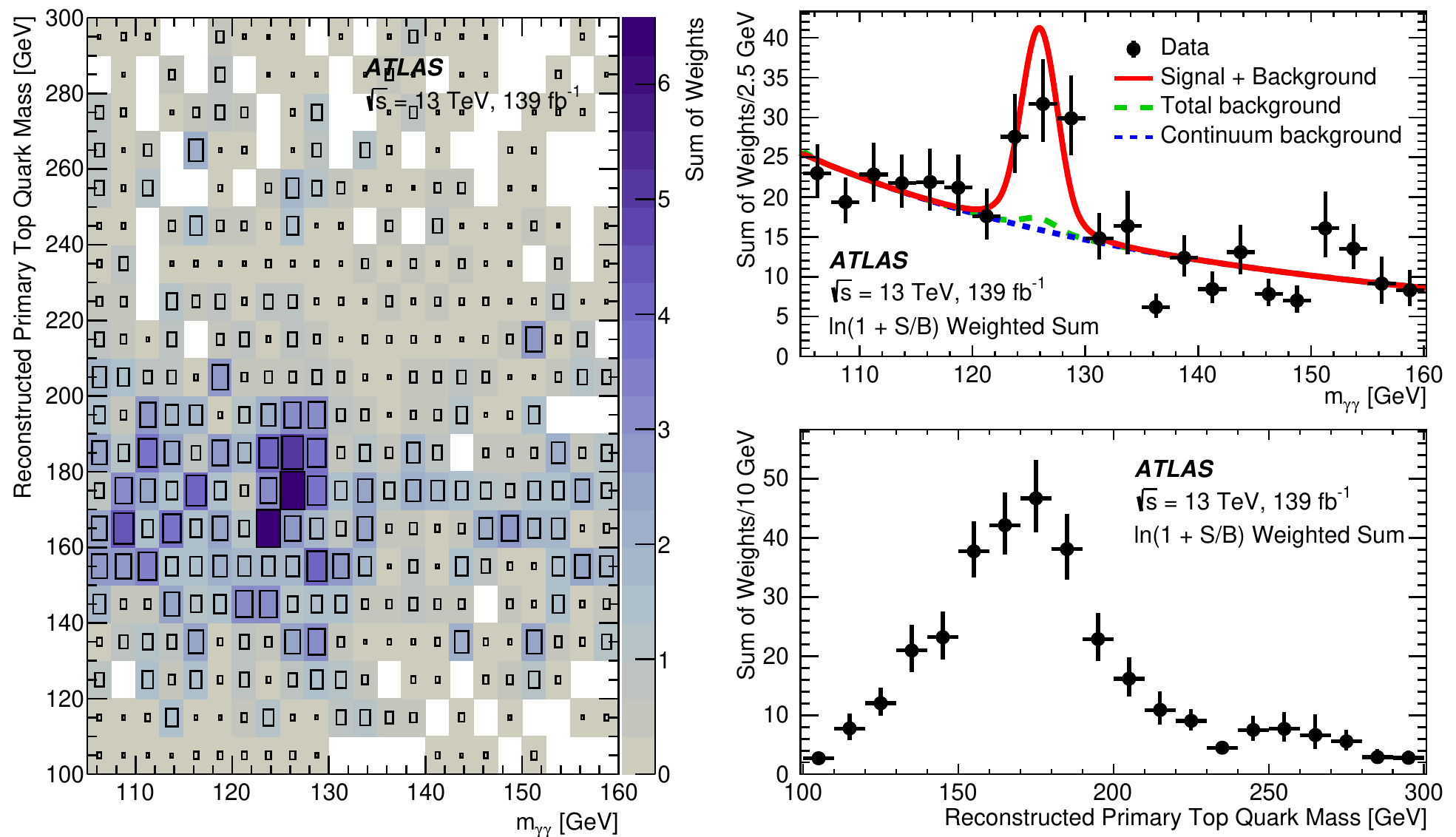}
\caption{Distribution of reconstructed primary top quark mass versus
reconstructed Higgs boson mass in the data events.
The right panels show the projections onto the Higgs boson mass and primary top quark mass axes.
In the upper panel, the fitted continuum background (blue),
the total background including non-$t\bar{t}H$/$tH$ Higgs boson production (green),
and the total fitted signal plus background (red) are shown.
The error bars on data are statistical.
}
\label{fig:mggmt_mc}
\end{figure}
 
Extraction of values for the top Yukawa coupling
requires additional information.
In particular, the BR of $H\to \gamma\gamma$ is needed to recover the
total Higgs boson production rate, and the Higgs boson coupling to gluons is needed
to account for the small ggF background.
The corresponding Higgs boson coupling
modifiers $\kappa_{\gamma}$ and $\kappa_{g}$ are measured in the
Run 2 Higgs boson coupling combination~\cite{HIGG-2018-57}.
This combination includes the first 80~\ifb\ of data used in this paper,
and $t\bar{t}H$ and $tH$ analyses from other decay channels.
The combination analysis is repeated without the $t\bar{t}H$ and $tH$ inputs and this result is
used to constrain $\kappa_{g}$ and $\kappa_{\gamma}$.
The impact on $\kappa_{g}$ and $\kappa_{\gamma}$
of removing input $t\bar{t}H$ and $tH$ analyses from the combination
is small.
Correlation of the systematic uncertainties between
the Higgs boson coupling combination and this analysis
is neglected.
The correlation has a small impact on $\alpha$, and a similar
effect on $\kappa_{t}$ as on signal strength reported in Ref.~\cite{HIGG-2018-57}.
This analysis is insensitive to the potential modifications of
ggF kinematics due to $CP$-mixing, which is therefore neglected.
The results of the fit for
$\kappa_t \cos(\alpha)$ and $\kappa_t \sin(\alpha)$
are shown  as contours in Figure~\ref{fig:s3:contours}.
A limit on $\alpha$ is set
without prior constraint on 
$\kappa_{t}$ in the fit:
$|\alpha| > 43^\circ$ is excluded at $95\%$ CL. The expected exclusion is $|\alpha| > 63^\circ$
under the $CP$-even hypothesis.
A value of $\alpha = 90~(180)^\circ$ is excluded at 3.9$\sigma$~(2.5$\sigma$).
A comparable study from the CMS experiment excluded
$\alpha = 90^\circ$ at 3.2$\sigma$~\cite{CMS-TTH-20-028}.
If $\kappa_{\gamma}$ and $\kappa_{g}$ are parameterized
using $\alpha$ and $\kappa_{t}$~\cite{Ellis:2013yxa},
the observed (expected) exclusion is $|\alpha| > 43~(56)^\circ$
without prior constraint on 
$\kappa_{t}$ in the fit.
The impact of the systematic uncertainties is negligible.

\begin{figure}[htbp]
\centering
\includegraphics[width=0.4\textwidth]{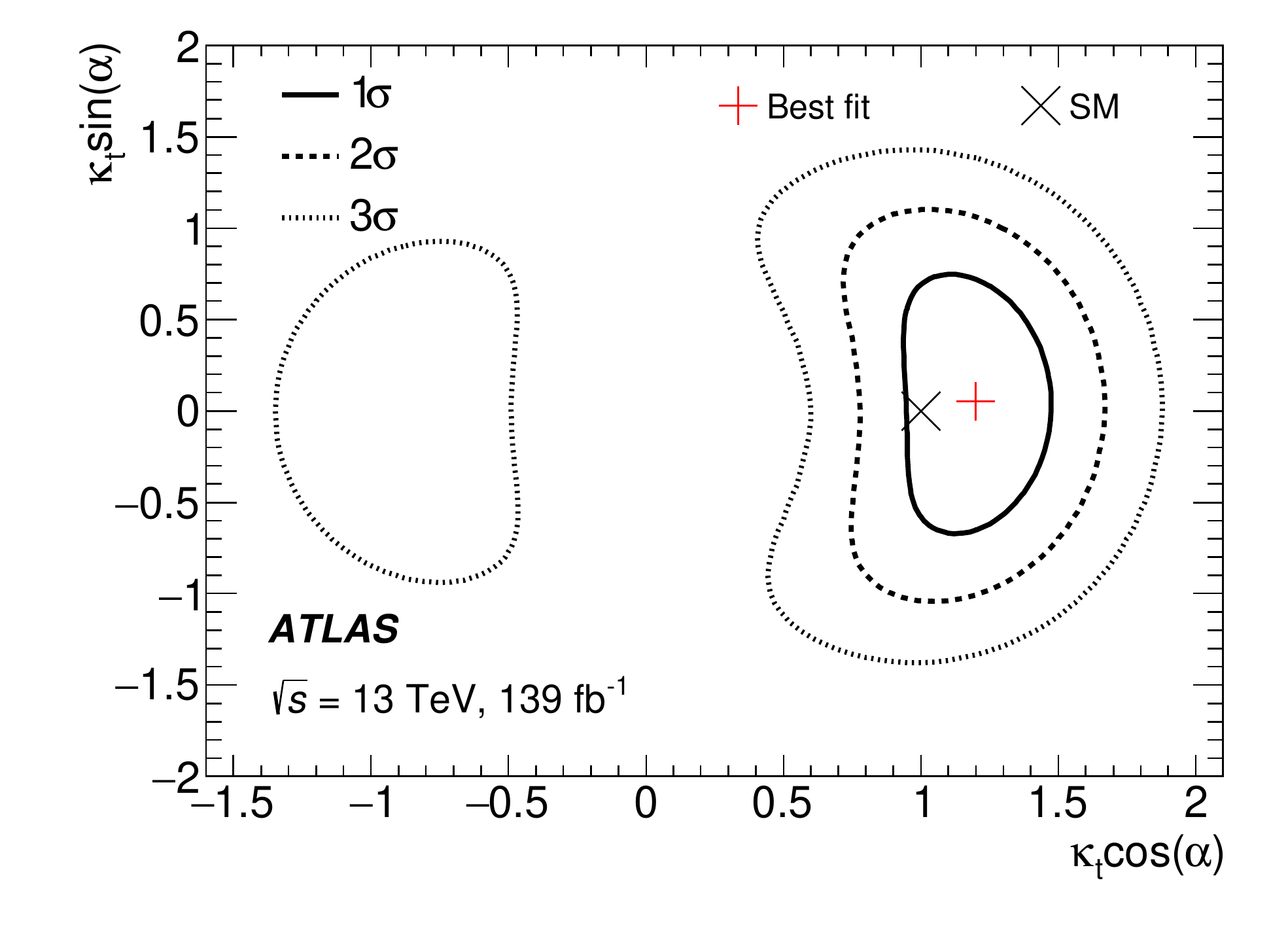}
\caption{Two-dimensional likelihood contours for $\kappa_{t}\cos(\alpha)$
and $\kappa_{t}\sin(\alpha)$
with ggF and \hyy\ constrained by the Higgs boson coupling combination.}
\label{fig:s3:contours}
\end{figure}
 
In summary, the production rate of
the Higgs boson in association with top quarks is measured and
the $CP$ property of the top Yukawa coupling is studied.
The no-$\ttH$ hypothesis is rejected with a significance of $5.2\sigma$,
and the measured $\sigma_{t\bar{t}H}\times B_{\gamma\gamma}$ is
$1.64\;^{+0.38}_{-0.36}\;\text{(stat.)}\;^{+0.17}_{-0.14}\;(\text{sys.})$ fb.
The measured rate for $\ttH$ is $1.43\;^{+0.33}_{-0.31}\;\text{(stat.)}\;^{+0.21}_{-0.15}\;(\text{sys.})$
times the SM expectation.
The $tH$ process is not observed and an upper limit of 12 times the SM expectation
is set on its rate at 95\% CL.
All measurements are consistent with the SM expectations and the possibility of $CP$-odd
couplings between the Higgs boson and top quark is severely constrained.
A pure $CP$-odd coupling is excluded at $3.9\sigma$,
and $|\alpha| > 43^\circ$ is excluded at $95\%$ CL.


We thank CERN for the very successful operation of the LHC, as well as the
support staff from our institutions without whom ATLAS could not be
operated efficiently.
 
We acknowledge the support of ANPCyT, Argentina; YerPhI, Armenia; ARC, Australia; BMWFW and FWF, Austria; ANAS, Azerbaijan; SSTC, Belarus; CNPq and FAPESP, Brazil; NSERC, NRC and CFI, Canada; CERN; CONICYT, Chile; CAS, MOST and NSFC, China; COLCIENCIAS, Colombia; MSMT CR, MPO CR and VSC CR, Czech Republic; DNRF and DNSRC, Denmark; IN2P3-CNRS and CEA-DRF/IRFU, France; SRNSFG, Georgia; BMBF, HGF and MPG, Germany; GSRT, Greece; RGC and Hong Kong SAR, China; ISF and Benoziyo Center, Israel; INFN, Italy; MEXT and JSPS, Japan; CNRST, Morocco; NWO, Netherlands; RCN, Norway; MNiSW and NCN, Poland; FCT, Portugal; MNE/IFA, Romania; MES of Russia and NRC KI, Russia Federation; JINR; MESTD, Serbia; MSSR, Slovakia; ARRS and MIZ\v{S}, Slovenia; DST/NRF, South Africa; MINECO, Spain; SRC and Wallenberg Foundation, Sweden; SERI, SNSF and Cantons of Bern and Geneva, Switzerland; MOST, Taiwan; TAEK, Turkey; STFC, United Kingdom; DOE and NSF, United States of America. In addition, individual groups and members have received support from BCKDF, CANARIE, Compute Canada and CRC, Canada; ERC, ERDF, Horizon 2020, Marie Sk{\l}odowska-Curie Actions and COST, European Union; Investissements d'Avenir Labex, Investissements d'Avenir Idex and ANR, France; DFG and AvH Foundation, Germany; Herakleitos, Thales and Aristeia programmes co-financed by EU-ESF and the Greek NSRF, Greece; BSF-NSF and GIF, Israel; CERCA Programme Generalitat de Catalunya and PROMETEO Programme Generalitat Valenciana, Spain; G\"{o}ran Gustafssons Stiftelse, Sweden; The Royal Society and Leverhulme Trust, United Kingdom.
 
The crucial computing support from all WLCG partners is acknowledged gratefully, in particular from CERN, the ATLAS Tier-1 facilities at TRIUMF (Canada), NDGF (Denmark, Norway, Sweden), CC-IN2P3 (France), KIT/GridKA (Germany), INFN-CNAF (Italy), NL-T1 (Netherlands), PIC (Spain), ASGC (Taiwan), RAL (UK) and BNL (USA), the Tier-2 facilities worldwide and large non-WLCG resource providers. Major contributors of computing resources are listed in Ref.~\cite{ATL-GEN-PUB-2016-002}.
 

\clearpage

\printbibliography

@article{LUCID2,
  author={G. Avoni and others},
  title={The new LUCID-2 detector for luminosity measurement and monitoring in ATLAS},
  journal={JINST},
  volume={13},
  number={07},
  pages={P07017},
  doi="10.1088/1748-0221/13/07/P07017",
  year={2018},
}

@Article{Alwall:2014hca,
      author         = "J. Alwall and others",
      title          = "{The automated computation of tree-level and
                        next-to-leading order differential cross sections, and
                        their matching to parton shower simulations}",
      journal        = "JHEP",
      volume         = "07",
      pages          = "079",
      doi            = "10.1007/JHEP07(2014)079",
      year           = "2014",
      eprint         = "1405.0301",
      archivePrefix  = "arXiv",
      primaryClass   = "hep-ph"
}

@article{Demartin:2014fia,
      author         = "Demartin, Federico and Maltoni, Fabio and Mawatari,
                        Kentarou and Page, Ben and Zaro, Marco",
      title          = "{Higgs characterisation at NLO in QCD: CP properties of
                        the top-quark Yukawa interaction}",
      journal        = "Eur. Phys. J. C",
      volume         = "74",
      year           = "2014",
      number         = "9",
      pages          = "3065",
      doi            = "10.1140/epjc/s10052-014-3065-2",
      eprint         = "1407.5089",
      archivePrefix  = "arXiv",
      primaryClass   = "hep-ph",
      reportNumber   = "CP3-14-59, LPN14-096, MCNET-14-21",
      SLACcitation   = "%%CITATION = ARXIV:1407.5089;%%"
}

@article{AmorDosSantos:2017ayi,
      author         = "Amor Dos Santos, S. and others",
      title          = "{Probing the CP nature of the Higgs coupling in $t{\bar
                        t}h$ events at the LHC}",
      journal        = "Phys. Rev. D",
      volume         = "96",
      year           = "2017",
      number         = "1",
      pages          = "013004",
      doi            = "10.1103/PhysRevD.96.013004",
      eprint         = "1704.03565",
      archivePrefix  = "arXiv",
      primaryClass   = "hep-ph",
      SLACcitation   = "%%CITATION = ARXIV:1704.03565;%%"
}

@article{Ellis:2013yxa,
      author         = "Ellis, John and Hwang, Dae Sung and Sakurai, Kazuki and
                        Takeuchi, Michihisa",
      title          = "{Disentangling Higgs-top couplings in associated
                        production}",
      journal        = "JHEP",
      volume         = "04",
      year           = "2014",
      pages          = "004",
      doi            = "10.1007/JHEP04(2014)004",
      eprint         = "1312.5736",
      archivePrefix  = "arXiv",
      primaryClass   = "hep-ph",
      reportNumber   = "KCL-PH-TH-2013-47, LCTS-2013-35, CERN-PH-TH-2013-312",
      SLACcitation   = "%%CITATION = ARXIV:1312.5736;%%"
}

@Article{Read:2002hq,
    Author = {Read, Alexander L.},
    Journal = {J. Phys. G},
    Pages = {2693},
    doi = "10.1088/0954-3899/28/10/313",
    Title = {{Presentation of search results: the \(CL_S\) technique}},
    Volume = {28},
    Year = {2002}
}

@article{Boudjema:2015nda,
      author         = "Boudjema, Fawzi and Godbole, Rohini M. and Guadagnoli,
                        Diego and Mohan, Kirtimaan A.",
      title          = "{Laboratory-frame observables for probing the top-Higgs boson
                        interaction}",
      journal        = "Phys. Rev. D",
      volume         = "92",
      year           = "2015",
      number         = "1",
      pages          = "015019",
      doi            = "10.1103/PhysRevD.92.015019",
      eprint         = "1501.03157",
      archivePrefix  = "arXiv",
      primaryClass   = "hep-ph",
      reportNumber   = "LAPTH-004-15, MSUHEP-150113",
      SLACcitation   = "%%CITATION = ARXIV:1501.03157;%%"
}

@article{efttop:2011prd,
      author         = "Cen Zhang and Scott Willenbrock",
      title          = "{Effective-field-theory approach to top-quark production and decay}",
      journal        = "Phys. Rev. D",
      volume         = "83",
      year           = "2011",
      number         = "3",
      pages          = "034006",
      doi            = "10.1103/PhysRevD.83.034006",
      eprint         = "1008.3869",
      archivePrefix  = "arXiv",
      primaryClass   = "hep-ph",
      reportNumber   = "LAPTH-004-15, MSUHEP-150113",
      SLACcitation   = "%%CITATION = ARXIV:1008.3869;%%"
}

@Article{Cacciari:2008gp,
     author    = "Cacciari, Matteo and Salam, Gavin P. and Soyez, Gregory",
     title     = "{The anti-\(k_t\) jet clustering algorithm}",
     journal   = "JHEP",
     volume    = "04",
     year      = "2008",
     pages     = "063",
     eprint    = "0802.1189",
     archivePrefix = "arXiv",
     primaryClass  =  "hep-ph",
     doi       = "10.1088/1126-6708/2008/04/063",
     SLACcitation  = "%%CITATION = 0802.1189;%%"
}

@Booklet{XGBoost,
  author    = {Tianqi Chen and
               Carlos Guestrin},
  title     = {XGBoost: {A} Scalable Tree Boosting System},
  eprint         =      "1603.02754",
  archivePrefix  =      "arXiv",
  primaryClass   = "cs.LG",
  journal   = {},
  volume    = {},
  doi       = {10.1145/2939672.2939785},
  bibsource = {dblp computer science bibliography, http://dblp.org}
}

@Article{Cowan:2010js, 
	Author = {Cowan, Glen and Cranmer, Kyle and Gross, Eilam and Vitells, Ofer}, 
	Eprint = {1007.1727}, 
	Archiveprefix = {arXiv}, 
	Primaryclass = {physics.data-an}, 
	Journal = {Eur. Phys. J. C}, 
	Pages = {1554}, 
    	doi = "10.1140/epjc/s10052-011-1554-0", 
	Title = {{Asymptotic formulae for likelihood-based tests of new physics}}, 
	Volume = {71}, 
	Year = {2011}, 
    	Addendum = "Erratum: \href{http://dx.doi.org/10.1140/epjc/s10052-013-2501-z}{Eur. Phys. J. C {\textbf{73}} (2013) 2501}" 
}

@techreport{YR4,
      author        = "{LHC Higgs Cross Section Working Group}",
      title         = "{Handbook of LHC Higgs Cross Sections: 4. Deciphering the Nature of the Higgs Sector}",
      institution   = "FERMILAB/CERN",
      reportNumber        = "FERMILAB-FN-1025-T/CERN-2017-002-M",
      year          = "2017",
      url           = "https://cds.cern.ch/record/2227475",
}

@article{Anastasiou:2015ema,
      author         = "Anastasiou, Charalampos and Duhr, Claude and Dulat, Falko
                        and Herzog, Franz and Mistlberger, Bernhard",
      title          = "{Higgs Boson Gluon-Fusion Production in QCD at Three
                        Loops}",
      journal        = "Phys.~Rev.~Lett.",
      volume         = "114",
      year           = "2015",
      pages          = "212001",
      doi            = "10.1103/PhysRevLett.114.212001",
      eprint         = "1503.06056",
      archivePrefix  = "arXiv",
      primaryClass   = "hep-ph",
      reportNumber   = "CERN-PH-TH-2015-055, CP3-15-07",
      SLACcitation   = "%%CITATION = ARXIV:1503.06056;%%"
}

@article{Buckley:2016prl,
          title = "{Boosting the Direct $CP$ Measurement of the Higgs-Top Coupling}",
          author = "Buckley, Matthew R. and Gon\ifmmode \mbox{\c{c}}\else \c{c}\fi{}alves, Dorival",
          journal = "Phys.~Rev.~Lett.",
          volume = "116",
          year = "2016",
          pages = "091801",
          doi = "10.1103/PhysRevLett.116.091801",
          eprint = "1507.07926", 
          archivePrefix  = "arXiv",
          primaryClass   = "hep-ph",
          reportNumber   = "IPPP-15-49, DCPT-15-98",
          SLACcitation   = "%%CITATION = ARXIV:1507.07926;%%"
}

@article{Anastasiou:2016cez,
      author         = "Anastasiou, Charalampos and Duhr, Claude and Dulat, Falko
                        and Furlan, Elisabetta and Gehrmann, Thomas and Herzog,
                        Franz and Lazopoulos, Achilleas and Mistlberger, Bernhard",
      title          = "{High precision determination of the gluon fusion Higgs
                        boson cross-section at the LHC}",
      journal        = "JHEP",
      volume         = "05",
      year           = "2016",
      pages          = "058",
      doi            = "10.1007/JHEP05(2016)058",
      eprint         = "1602.00695",
      archivePrefix  = "arXiv",
      primaryClass   = "hep-ph",
      reportNumber   = "CP3-16-01, ZU-TH-27-15, NIKHEF-2016-004,
                        CERN-TH-2016-006",
      SLACcitation   = "%%CITATION = ARXIV:1602.00695;%%"
}

@article{Actis:2008ug,
      author         = "Actis, Stefano and Passarino, Giampiero and Sturm,
                        Christian and Uccirati, Sandro",
      title          = "{NLO electroweak corrections to Higgs boson production at
                        hadron colliders}",
      journal        = "Phys.~Lett.~B",
      volume         = "670",
      pages          = "12-17",
      doi            = "10.1016/j.physletb.2008.10.018",
      year           = "2008",
      eprint         = "0809.1301",
      archivePrefix  = "arXiv",
      primaryClass   = "hep-ph",
      reportNumber   = "PITHA-08-20, SFB-CPP-08-62, TTP08-38",
      SLACcitation   = "%%CITATION = ARXIV:0809.1301;%%",
}

@article{Anastasiou:2008tj,
      author         = "Anastasiou, Charalampos and Boughezal, Radja and
                        Petriello, Frank",
      title          = "{Mixed QCD-electroweak corrections to Higgs boson
                        production in gluon fusion}",
      journal        = "JHEP",
      volume         = "04",
      pages          = "003",
      doi            = "10.1088/1126-6708/2009/04/003",
      year           = "2009",
      eprint         = "0811.3458",
      archivePrefix  = "arXiv",
      primaryClass   = "hep-ph",
      SLACcitation   = "%%CITATION = ARXIV:0811.3458;%%",
}

@article{Ciccolini:2007jr,
      author         = "Ciccolini, M. and Denner, Ansgar and Dittmaier, S.",
      title          = "{Strong and Electroweak Corrections to the Production of a
                        Higgs Boson + 2 Jets via Weak Interactions at the Large Hadron Collider}",
      journal        = "Phys.~Rev.~Lett.",
      volume         = "99",
      pages          = "161803",
      doi            = "10.1103/PhysRevLett.99.161803",
      year           = "2007",
      eprint         = "0707.0381",
      archivePrefix  = "arXiv",
      primaryClass   = "hep-ph",
      reportNumber   = "MPP-2007-82, PSI-PR-07-03",
      SLACcitation   = "%%CITATION = ARXIV:0707.0381;%%",                                                 
}

@article{Ciccolini:2007ec,
      author         = "Ciccolini, Mariano and Denner, Ansgar and Dittmaier, Stefan",
      title          = "{Electroweak and QCD corrections to Higgs production via
                        vector-boson fusion at the CERN LHC}",
      journal        = "Phys.~Rev.~D",
      volume         = "77",
      pages          = "013002",
      doi            = "10.1103/PhysRevD.77.013002",
      year           = "2008",
      eprint         = "0710.4749",
      archivePrefix  = "arXiv",
      primaryClass   = "hep-ph",
      reportNumber   = "MPP-2007-152, PSI-PR-07-06, UWTHPH-2007-26",
      SLACcitation   = "%%CITATION = ARXIV:0710.4749;%%",                                                 
}

@article{Bolzoni:2010xr,
      author         = "Bolzoni, Paolo and Maltoni, Fabio and Moch, Sven-Olaf and
                        Zaro, Marco",
      title          = "{Higgs Boson Production via Vector-Boson Fusion at Next-to-Next-to-Leading Order in QCD}",
      journal        = "Phys.~Rev.~Lett.",
      volume         = "105",
      pages          = "011801",
      doi            = "10.1103/PhysRevLett.105.011801",
      year           = "2010",
      eprint         = "1003.4451",
      archivePrefix  = "arXiv",
      primaryClass   = "hep-ph",
      reportNumber   = "DESY-10-042, SFB-CPP-10-23, CP3-10-12",
      SLACcitation   = "%%CITATION = ARXIV:1003.4451;%%",                                                 
}

@article{Brein:2003wg,
      author         = "Brein, Oliver and Djouadi, Abdelhak and Harlander,
                        Robert",
      title          = "{NNLO QCD corrections to the Higgs-strahlung processes at
                        hadron colliders}",
      journal        = "Phys.~Lett.~B",
      volume         = "579",
      pages          = "149-156",
      doi            = "10.1016/j.physletb.2003.10.112",
      year           = "2004",
      eprint         = "hep-ph/0307206",
      archivePrefix  = "arXiv",
      reportNumber   = "MPP-2003-35, CERN-TH-2003-161, PM-03-16",
      SLACcitation   = "%%CITATION = HEP-PH/0307206;%%",                                                  
}

@article{Altenkamp:2012sx,
      author            = "Altenkamp, Lukas and Dittmaier, Stefan and Harlander, Robert V. and Rzehak, Heidi and Zirke, Tom J. E.",
      title             = "{Gluon-induced Higgs-strahlung at next-to-leading order QCD}",
      journal           = "JHEP",
      volume            = "02",
      year              = "2013",
      pages             = "078",
      doi               = "10.1007/JHEP02(2013)078",
      eprint            = "1211.5015",
      archivePrefix     = "arXiv",
      primaryClass      = "hep-ph",
      reportNumber      = "CERN-PH-TH-2012-312, FR-PHENO-2012-023, WUB-12-21",
      SLACcitation      = "%%CITATION = ARXIV:1211.5015;%%"                                               
}

@article{Denner:2011id,
      author                    = "Denner, Ansgar and Dittmaier, Stefan and Kallweit, Stefan and M{\"u}ck, Alexander",
      title                     = "{Electroweak corrections to Higgs-strahlung off W/Z bosons at the Tevatron and the LHC with HAWK}",
      journal                   = "JHEP",
      volume            = "03",
      pages             = "075",
      doi                       = "10.1007/JHEP03(2012)075",
      year                      = "2012",
      eprint                    = "1112.5142",
      archivePrefix     = "arXiv",
      primaryClass      = "hep-ph",
      reportNumber      = "FR-PHENO-2011-025, PSI-PR-11-04, ZU-TH-29-11, TTK-11-61",
      SLACcitation      = "%%CITATION = ARXIV:1112.5142;%%",                                              
}

@article{Beenakker:2002nc,
      author         = "Beenakker, W. and others",
      title          = "{NLO QCD corrections to $t\bar{t} H$ production in hadron
                        collisions}",
      journal        = "Nucl.~Phys.~B",
      volume         = "653",
      pages          = "151-203",
      doi            = "10.1016/S0550-3213(03)00044-0",
      year           = "2003",
      eprint         = "hep-ph/0211352",
      archivePrefix  = "arXiv",
      reportNumber   = "DESY-02-177, EDINBURGH-2002-18, MPI-PHT-2002-70,
                        PSI-PR-02-22",
      SLACcitation   = "%%CITATION = HEP-PH/0211352;%%",                                                  
}

@article{Dawson:2003zu,
      author         = "Dawson, S. and Jackson, C. and Orr, L.H. and Reina, L.
                        and Wackeroth, D.",
      title          = "{Associated Higgs boson production with top quarks at the CERN
                        Large Hadron Collider: NLO QCD corrections}",
      journal        = "Phys.~Rev.~D",
      volume         = "68",
      pages          = "034022",
      doi            = "10.1103/PhysRevD.68.034022",
      year           = "2003",
      eprint         = "hep-ph/0305087",
      archivePrefix  = "arXiv",
      reportNumber   = "BNL-HET-03-9, FSU-HEP-2003-0503, UB-HET-03-02",
      SLACcitation   = "%%CITATION = HEP-PH/0305087;%%",                                                  
}

@article{Yu:2014cka,
      author                    = "Zhang, Yu and Ma, Wen-Gan and Zhang, Ren-You and Chen, Chong and Guo, Lei",
      title                     = "{QCD NLO and EW NLO corrections to $t\bar{t}H$ production with top quark decays at hadron collider}",
      journal                   = "Phys.~Lett.~B",
      volume            = "738",
      year                      = "2014",
      pages             = "1-5",
      doi                       = "10.1016/j.physletb.2014.09.022",
      eprint                    = "1407.1110",
      archivePrefix     = "arXiv",
      primaryClass      = "hep-ph",
      SLACcitation      = "%%CITATION = ARXIV:1407.1110;%%"                                               
}

@article{Frixione:2015zaa,
      author         = "Frixione, S. and Hirschi, V. and Pagani, D. and Shao, H.-S. and Zaro, M.",
      title          = "{Electroweak and QCD corrections to top-pair
                        hadroproduction in association with heavy bosons}",
      journal        = "JHEP",
      volume         = "06",
      year           = "2015",
      pages          = "184",
      doi            = "10.1007/JHEP06(2015)184",
      eprint         = "1504.03446",
      archivePrefix  = "arXiv",
      primaryClass   = "hep-ph",
      reportNumber   = "CERN-PH-TH-2015-083, CP3-15-10, LPN15-022,
                        SLAC-PUB-16253",
      SLACcitation   = "%%CITATION = ARXIV:1504.03446;%%"
}

@article{Dawson:2003kb,
        author          = "Dawson, S. and Jackson, C.B. and Reina, L. and Wackeroth, D.",
        title                   = "{Exclusive Higgs boson production with bottom quarks at hadron colliders}",
        journal         = "Phys.~Rev.~D",
        volume          = "69",
        pages           = "074027",
        doi                     = "10.1103/PhysRevD.69.074027",
        year            = "2004",
        eprint          = "hep-ph/0311067",
        archivePrefix   = "arXiv",
        reportNumber    = "BNL-HET-03-17, FSU-HEP-2003-1015, UB-HET-03-05",
        SLACcitation    = "%%CITATION = HEP-PH/0311067;%%",                                               
}

@article{Dittmaier:2003ej,
        author          = "Dittmaier, Stefan and {Kr\"amer}, Michael and Spira, Michael",
        title                   = "{Higgs radiation off bottom quarks at the Fermilab Tevatron and the CERN LHC}",
        journal         = "Phys.~Rev.~D",
        volume          = "70",
        pages           = "074010",
        doi                     = "10.1103/PhysRevD.70.074010",
        year            = "2004",
        eprint          = "hep-ph/0309204",
        archivePrefix   = "arXiv",
        reportNumber = "EDINBURGH-2003-13, MPP-2003-49, PSI-PR-03-14",
        SLACcitation    = "%%CITATION = HEP-PH/0309204;%%",                                               
}

@Booklet{Harlander:2011aa,
      author         = "Harlander, Robert and Kr{\"a}mer, Michael and Schumacher, Markus",
      title          = "{Bottom-quark associated Higgs-boson production:
                        reconciling the four- and five-flavour scheme approach}",
      year           = "2011",
      eprint         = "1112.3478",
      archivePrefix  = "arXiv",
      primaryClass   = "hep-ph",
      reportNumber   = "CERN-PH-TH-2011-134, FR-PHENO-2011-009, TTK-11-17,
                        WUB-11-04",
      SLACcitation   = "%%CITATION = ARXIV:1112.3478;%%"                                                  
}

@article{Demartin:2015uha,
      author                    = "Demartin, Federico and Maltoni, Fabio and Mawatari, Kentarou and Zaro, Marco",
      title                     = "{Higgs production in association with a single top quark at the LHC}",
      journal                   = "Eur.~Phys.~J.~C",
      volume            = "75",
      year                      = "2015",
      pages             = "267",
      doi                       = "10.1140/epjc/s10052-015-3475-9",
      eprint                    = "1504.00611",
      archivePrefix     = "arXiv",
      primaryClass      = "hep-ph",
      reportNumber      = "MCNET-15-07, CP3-15-08",
      SLACcitation      = "%%CITATION = ARXIV:1504.00611;%%"                                              
}

@article{Djouadi:1997yw,
        author          = "Djouadi, A. and Kalinowski, J. and Spira, M.",
        title                   = "{HDECAY: a program for Higgs boson decays in the Standard Model and its supersymmetric extension}",
        journal         = "Comput. Phys. Commun.",
        volume          = "108",
        pages           = "56-74",
        doi                     = "10.1016/S0010-4655(97)00123-9",
        year            = "1998",
        eprint          = "hep-ph/9704448",
        archivePrefix   = "arXiv",
        reportNumber    = "DESY-97-079, IFT-96-29, PM-97-04",
        SLACcitation    = "%%CITATION = HEP-PH/9704448;%%",                                               
}

@article{Djouadi:2006bz,
      author                    = "Djouadi, A. and M{\"u}hlleitner, M. M. and Spira, M.",
      title                     = "{Decays of supersymmetric particles: The Program SUSY-HIT (SUspect-SdecaY-Hdecay-InTerface)}",
      booktitle         = "{Physics at LHC. Proceedings, 3rd Conference, Cracow, Poland, July 3-8, 2006}",
      journal                   = "Acta Phys. Polon. B",
      volume            = "38",
      year                      = "2007",
      pages             = "635-644",
      eprint                    = "hep-ph/0609292",
      archivePrefix     = "arXiv",
      reportNumber      = "CERN-PH-TH-2006-200",
      SLACcitation      = "%%CITATION = HEP-PH/0609292;%%"                                                
}

@Article{CMS-TTH-20-028,
    author         = "{CMS Collaboration}",
    title          = "{Measurements of \(t\bar{t}H\) Production and the $CP$ Structure of the Yukawa Interaction between the Higgs Boson and Top Quark in the Diphoton Decay Channel}",
    journal        = "Phys. Rev. Lett.",
    volume         = "125",
    year           = "2020",
    pages          = "061801",
    doi            = "10.1103/PhysRevLett.125.061801",
    reportNumber   = "CERN-EP-2020-028",
    eprint         = "2003.10866",
    archivePrefix  = "arXiv",
    primaryClass   = "hep-ex",
}

@article{Bredenstein:2006ha,
      author         = "Bredenstein, A. and Denner, Ansgar and Dittmaier, S. and
                        Weber, M. M.",
      title          = "{Radiative corrections to the semileptonic and hadronic
                        Higgs-boson decays $H \to  WW / ZZ \to 4$ fermions}",
      journal        = "JHEP",
      volume         = "02",
      year           = "2007",
      pages          = "080",
      doi            = "10.1088/1126-6708/2007/02/080",
      eprint         = "hep-ph/0611234",
      archivePrefix  = "arXiv",
      reportNumber   = "KEK-CP-187, KEK-PREPRINT-2006-47, MPP-2006-138,
                        PSI-PR-06-11, WUB-06-08",
      SLACcitation   = "%%CITATION = HEP-PH/0611234;%%"                                                   
}

@article{Bredenstein:2006rh,
      author         = "Bredenstein, A. and Denner, Ansgar and Dittmaier, S. and
                        Weber, M. M.",
      title          = "{Precise predictions for the Higgs-boson decay $H\to WW/ZZ \to 4$ leptons}",
      journal        = "Phys.~Rev.~D",
      volume         = "74",
      year           = "2006",
      pages          = "013004",
      doi            = "10.1103/PhysRevD.74.013004",
      eprint         = "hep-ph/0604011",
      archivePrefix  = "arXiv",
      reportNumber   = "MPP-2005-24, PSI-PR-06-05",
      SLACcitation   = "%%CITATION = HEP-PH/0604011;%%"                                                   
}

@misc{Goodnessfit,
      title         = "{Generalization of Chisquare Goodness-of-Fit Test for 
                       Binned Data Using Saturated Models, with Application to Histograms}",
     author = "{R. D. Cousins}",
     howpublished = {\url{http://www.physics.ucla.edu/~cousins/stats/cousins_saturated.pdf}},
     year = 2013
}

@article{Ball:2012cx,
      author         = "Ball, Richard D. and others",
      title          = "{Parton distributions with LHC data}",
      journal        = "Nucl. Phys. B",
      volume         = "867",
      year           = "2013",
      pages          = "244",
      doi            = "10.1016/j.nuclphysb.2012.10.003",
      eprint         = "1207.1303",
      archivePrefix  = "arXiv",
      primaryClass   = "hep-ph",
      reportNumber   = "EDINBURGH-2012-08, IFUM-FT-997, FR-PHENO-2012-014,
                        RWTH-TTK-12-25, CERN-PH-TH-2012-037, SFB-CPP-12-47\,
                        --CERN-PH-TH-2012-037",
      SLACcitation   = "%%CITATION = ARXIV:1207.1303;%%"
}

@article{Ball:2014uwa,
      author         = "Ball, Richard D. and others",
      title          = "{Parton distributions for the LHC run II}",
      collaboration  = "NNPDF",
      journal        = "JHEP",
      volume         = "04",
      year           = "2015",
      pages          = "040",
      doi            = "10.1007/JHEP04(2015)040",
      eprint         = "1410.8849",
      archivePrefix  = "arXiv",
      primaryClass   = "hep-ph",
      reportNumber   = "EDINBURGH-2014-15, IFUM-1034-FT, CERN-PH-TH-2013-253,
                        OUTP-14-11P, CAVENDISH-HEP-14-11",
      SLACcitation   = "%%CITATION = ARXIV:1410.8849;%%"
}

@Article{Pumplin:2002vw,
     author    = "Pumplin, J. and others",
     title     = "{New Generation of Parton Distributions with Uncertainties from Global QCD Analysis}",
     journal   = "JHEP",
     volume    = "07",
     year      = "2002",
     pages     = "012",
     doi       = "10.1088/1126-6708/2002/07/012",
     eprint    = "hep-ph/0201195",
     archivePrefix = "arXiv",
     SLACcitation  = "%%CITATION = HEP-PH/0201195;%%"
}

@article{Geant4,
      author         = "Agostinelli, S. and others",
      title          = "{\textsc{Geant4} -- a simulation toolkit}",
      journal        = "Nucl. Instrum. Meth. A",
      volume         = "506",
      year           = "2003",
      pages          = "250",
      doi            = "10.1016/S0168-9002(03)01368-8",
      reportNumber   = "SLAC-PUB-9350, FERMILAB-PUB-03-339",
      SLACcitation   = "%%CITATION = NUIMA,A506,250;%%"
}

@article{Powheg_Alioli,
      Eprint = {1002.2581},
      Primaryclass = {hep-ph},
      Archiveprefix = {arXiv},
      Title = {{A general framework for implementing NLO calculations in shower Monte Carlo programs: the \textsc{Powheg} BOX}},
      Author = {Alioli, Simone and others},
      Doi = {10.1007/JHEP06(2010)043},
      Journal = {JHEP},
      Pages = {043},
      Reportnumber = {DESY-10-018, SFB-CPP-10-22, IPPP-10-11, DCPT-10-22},
      Slaccitation = {%%CITATION = ARXIV:1002.2581;%%},
      Volume = {06},
      Year = {2010}
}

@article{Pythia8,
      author         = "{T. Sj\"ostrand, S. Mrenna and P. Skands}",
      title          = "{A brief introduction to PYTHIA 8.1}",
      journal        = "Comput. Phys. Commun.",
      volume         = "178",
      pages          = "852-867",
      doi            = "10.1016/j.cpc.2008.01.036",
      year           = "2008",
      eprint         = "0710.3820",
      archivePrefix  = "arXiv",
      primaryClass   = "hep-ph",
      reportNumber   = "CERN-LCGAPP-2007-04, LU-TP-07-28,
                        FERMILAB-PUB-07-512-CD-T",
      SLACcitation   = "%%CITATION = ARXIV:0710.3820;%%",
}

@article{Hamilton:2013fea,
      author         = "Hamilton, Keith and Nason, Paolo and Re, Emanuele and
                        Zanderighi, Giulia",
      title          = "{NNLOPS simulation of Higgs boson production}",
      journal        = "JHEP",
      volume         = "10",
      year           = "2013",
      pages          = "222",
      doi            = "10.1007/JHEP10(2013)222",
      eprint         = "1309.0017",
      archivePrefix  = "arXiv",
      primaryClass   = "hep-ph",
      reportNumber   = "MCNET-13-11, CERN-PH-TH-2013-205, OUTP-13-18P",
      SLACcitation   = "%%CITATION = ARXIV:1309.0017;%%"
}

@article{Nason:2010jhep,
      author        = "Nason, Paolo and Oleari, Carlo",
      title         = "{NLO Higgs boson production via vector-boson fusion matched with shower in POWHEG}",
      journal       = "JHEP",
      volume        = "02",
      year          = "2010",
      pages         = "037",
      eprint        = "0911.5299",
      archivePrefix = "arXiv",
      primaryClass  =  "hep-ph",
      doi     = "10.1007/JHEP02(2010)037",
      SLACcitation  = "%%CITATION = 0911.5299;%%"
}

\clearpage 
 
\begin{flushleft}
\hypersetup{urlcolor=black}
{\Large The ATLAS Collaboration}

\bigskip

\AtlasOrcid[0000-0002-6665-4934]{G.~Aad}$^\textrm{\scriptsize 102}$,    
\AtlasOrcid[0000-0002-5888-2734]{B.~Abbott}$^\textrm{\scriptsize 128}$,    
\AtlasOrcid{D.C.~Abbott}$^\textrm{\scriptsize 103}$,    
\AtlasOrcid[0000-0002-2788-3822]{A.~Abed~Abud}$^\textrm{\scriptsize 36}$,    
\AtlasOrcid[0000-0002-1002-1652]{K.~Abeling}$^\textrm{\scriptsize 53}$,    
\AtlasOrcid[0000-0002-2987-4006]{D.K.~Abhayasinghe}$^\textrm{\scriptsize 94}$,    
\AtlasOrcid[0000-0002-8496-9294]{S.H.~Abidi}$^\textrm{\scriptsize 166}$,    
\AtlasOrcid[0000-0002-8279-9324]{O.S.~AbouZeid}$^\textrm{\scriptsize 40}$,    
\AtlasOrcid{N.L.~Abraham}$^\textrm{\scriptsize 155}$,    
\AtlasOrcid[0000-0001-5329-6640]{H.~Abramowicz}$^\textrm{\scriptsize 160}$
% ...
\bigskip
\\

% ...
$^{36}$CERN, Geneva; Switzerland.\\
% ...
$^{40}$Niels Bohr Institute, University of Copenhagen, Copenhagen; Denmark.\\
% ...
$^{53}$II. Physikalisches Institut, Georg-August-Universit\"{a}t G\"ottingen, G\"ottingen; Germany.\\
% ...
$^{94}$Department of Physics, Royal Holloway University of London, Egham; United Kingdom.\\
% ...
$^{102}$CPPM, Aix-Marseille Universit\'e, CNRS/IN2P3, Marseille; France.\\
$^{103}$Department of Physics, University of Massachusetts, Amherst MA; United States of America.\\
% ...
$^{128}$Homer L. Dodge Department of Physics and Astronomy, University of Oklahoma, Norman OK; United States of America.\\
% ...
$^{155}$Department of Physics and Astronomy, University of Sussex, Brighton; United Kingdom.\\
% ...
$^{160}$Raymond and Beverly Sackler School of Physics and Astronomy, Tel Aviv University, Tel Aviv; Israel.\\
% ...
$^{166}$Department of Physics, University of Toronto, Toronto ON; Canada.\\
% ...

\end{flushleft}


\end{document}